\documentstyle[sprocl]{article}

\input epsf
\def\be{\begin{equation}}
\def\ee{\end{equation}}
\def\bea{\begin{eqnarray}}
\def\eea{\end{eqnarray}}

\begin{document}

\title{INFLATION IN \\
SOFTLY BROKEN SEIBERG-WITTEN MODELS}

\author{J. GARC\'IA-BELLIDO}

\address{TH-Division, C.E.R.N., CH-1211 Gen\`eve 23, Swirtzerland}

\maketitle\abstracts{In a recent paper we proposed a new model of
  inflation based on the soft-breaking of N=2 supersymmetric SU(2)
  Yang-Mills theory. The advantage of such a model is the fact that we
  can write an {\em exact} expression for the effective scalar
  potential, including perturbative and non-perturbative effects. We
  find that the scalar condensate that plays the role of the inflaton
  can drive a long period of cosmological expansion in the weak
  coupling Higgs region, and end inflation in the strong coupling
  monopole region, where reheating takes place. The model predicts the
  right amount of temperature anisotropies in the microwave
  background, a precise spectral tilt, $n=0.91$, and negligible
  gravitational wave perturbations.}

In a recent paper~\cite{JGB} we proposed a new model of inflation
where the role of the inflaton is played by the scalar condensate 
$u=Tr\,\varphi^2$ that
parametrizes inequivalent vacua in N=2 SU(2) super Yang-Mills. Duality
and analyticity arguments allow one to write down an {\em exact}
effective action for the light degrees of freedom in both the weak and
in the strong coupling regions~\cite{SW}. The soft breaking of N=2
directly down to N=0 via a spurion superfield preserves the
analyticity properties of the Seiberg-Witten solution and produces a
low energy effective scalar potential which includes all perturbative
and non-perturbative effects~\cite{Luis}. This powerful result was
studied in the context of low energy QCD. We simply realized that this
{\em exact} scalar potential could be consistently used at much higher
energies, of order the GUT scale, and be responsible for cosmological
inflation~\cite{JGB}. The advantage with respect to other inflationary
models based on supersymmetry is the complete control we have on
the scalar potential, both along the quasi-flat direction and in the
true vacuum of the theory, where reheating takes place.

In the Higgs region, along the positive real axis, it is possible to
write in a compact way the K\"ahler metric and the scalar potential 
as~\cite{JGB}
\begin{eqnarray}\label{kahler}
{\cal K}(u)&=&{2k^2\over\Lambda^2\pi^2}\,K K' \,,\\ \label{poten}
V(u)&=&{f_0^2\Lambda^2\over\pi^2}\left[1 - 2\Big({K-E\over k^2K}-
{1\over2}\Big)\right]\,.
\end{eqnarray}
The functions $K(k)$ and $E(k)$ are complete elliptic functions of the
first and second kind respectively, and $E'(k)\equiv E(k')$, where
$k'^2 + k^2 = 1$. All these are functions of $k^2=2/(1+u/\Lambda^2)$
in the complex moduli plane $u$. The scalar potential is thus a
non-trivial function of the inflaton field, and depends on only two
parameters, the dynamical scale $\Lambda$ and the supersymmetry
breaking scale $f_0$, satisfying $f_0<\Lambda<M\equiv M_{\rm P}/
\sqrt{8\pi}$ for the consistency of the theory~\cite{Luis}. 
In order to study the cosmological evolution of
the inflaton under the potential (\ref{poten}) one should embed this
model in supergravity. It is possible to show~\cite{Luis} that the
only gravitational corrections to the potential are proportional to
$m_{3/2}^2f_0^2$, and therefore completely negligible in our case,
since inflation in this model turns out to occur at the GUT scale, see
below, and thus much above the phenomenological gravitino mass scale.

One also has to take into account the non-trivial K\"ahler metric for
$u$~\cite{Luis}. The Lagrangian for the scalar field $u$ in a curved
background can then be written as
\begin{equation}\label{Lagrange}
{\cal L} = {1\over2}{\cal K}(u)\,g^{\mu\nu}\partial_\mu u\,
\partial_\nu\bar u - V(u)
= {1\over2} g^{\mu\nu}\partial_\mu \phi\,
\partial_\nu\bar \phi - V(\phi)\,,
\end{equation}
where $g_{\mu\nu}$ is the spacetime metric, and we have redefined the
inflaton field through $d\phi \equiv {\cal K}(u)^{1/2} du$. We have
plotted in Fig.~1 the scalar potential $V(\phi)$ as a function of the
real and imaginary parts of $\phi$. The vacuum energy, $V_0 =
f_0^2\Lambda^2/\pi^2$, was added to the scalar potential in order to
ensure that the absolute minimum is at zero cosmological constant.

\begin{figure}[t]
\centering
\hspace*{-4mm}
\leavevmode\epsfysize=4.25cm\epsfbox{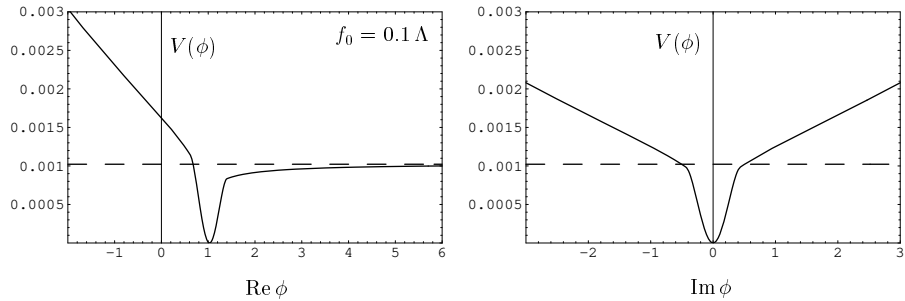}
\caption{\label{fig1} The exact scalar potential $V(\phi)$ in
  moduli space, for $f_0 = 0.1\Lambda$. The left panel shows the
  potential as a function of ${\rm Re}\,\phi$ for ${\rm Im}\,\phi=0$
  and presents a minimum at $\phi_{\rm min} = \Lambda + f_0/\sqrt8$.
  The right panel shows the potential as a function of ${\rm
    Im}\,\phi$ for ${\rm Re}\,\phi = \phi_{\rm min}$. The plateau
  (dashed line) corresponds to a vacuum energy density of $V_0 =
  f_0^2\Lambda^2/\pi^2$.}
\end{figure}

The flatness of this potential at ${\rm Im}\,\phi =0, {\rm Re}\,\phi >
\phi_{\rm min}$ looks like an excellent candidate for inflation. It
was not included by hand but arised naturally from the soft-breaking
of supersymmetry~\cite{Luis}, albeit with a complicated functional
form (\ref{poten}). There are two parameters in this model, $\Lambda$
and $f_0$. Non of these have to be fine tuned to be small in order to
have successful inflation.  Moreover, the trajectories away from the
positive real axis do not give inflation~\cite{JGB}. 

We will now consider the range of values of $f_0$ and $\Lambda$ that
give a phenomenologically viable model, applying the usual machinery
to study inflationary cosmology with a scalar field
potential~\cite{book,LL93}. It turns out that for all values of the
parameters $\Lambda$ and $f_0<\Lambda$, the extreme flatness of the
potential at $\phi> \phi_{\rm min}$ allows one to use the slow roll
approximation~\cite{LL93} in the Higgs region all the way to the
monopole region, where the slope of the potential is so large that
inflation ends and reheating starts as the condensate oscillates
around the minimum. As a consequence of the factorization of the
symmetry breaking parameter $f_0^2$ in the potential (\ref{poten}),
the slow-roll parameters do not depend on $f_0$~\cite{JGB}. This
further simplifies our analysis. For a given value of $\Lambda/M$ it
is easy to find the value of $\phi_e$ at the end of inflation and from
there compute the value $\phi_{60}$ corresponding to $N=60$ $e$-folds
from the end of inflation. 

Quantum fluctuations of the scalar condensate, $\delta \phi$, will
create perturbations in the metric, ${\cal R} = H\delta\phi/\dot\phi$,
which cross the Hubble scale during inflation and later re-enter
during the matter era. Those fluctuations corresponding to the scale
of the present horizon left 60 $e$-folds before the end of inflation,
and are responsible via the Sachs-Wolfe effect for the observed
temperature anisotropies in the microwave background~\cite{COBE}. From
the amplitude and spectral tilt of these temperature fluctuations we
can constrain the values of the parameters $\Lambda$ and $f_0$.

Present observations of the power spectrum of temperature anisotropies
on various scales, from COBE DMR to Saskatoon and CAT experiments,
impose the following constraints on the amplitude of the tenth
multipole and the tilt of the spectrum~\cite{Charley},
\begin{eqnarray}\label{ampli}
Q_{10} &=& 17.5\pm1.1\ \mu{\rm K}\,,\\
n &=& 0.91 \pm 0.10\,.\label{tilt}
\end{eqnarray}
Assuming that the dominant contribution to the CMB anisotropies comes
form the scalar metric perturbations, we can write~\cite{Charley} $A_S
= 5\times10^{-5}\,(Q_{10}/17.6\ \mu{\rm K})$, where $A_S^2 = (H/2\pi
M)^2/2\epsilon$ and $n = 1 + 2\eta - 6\epsilon$, in the slow-roll
approximation~\cite{LL93}. Since during inflation at large values of
$\phi$, corresponding to $N=60$, the rate of expansion is dominated by
the vacuum energy density $V_0$, we can write, to very good
approximation, $H^2=f_0^2\Lambda^2/3\pi^2M^2$, and thus the amplitude
of scalar metric perturbations is
\begin{eqnarray}\label{PR}
A_S^2 = {1\over24\pi^4}{f_0^2\Lambda^2\over M^4}
{1\over\epsilon_{60}}\,.
\end{eqnarray}

Let us consider, for example, a model with $\Lambda=0.1M$. In that
case the end of inflation occurs at $\phi_e\simeq1.5\Lambda$, still in
the Higgs region, and 60 $e$-folds correspond to a relatively large
value, $\phi_{60} = 14\Lambda$, deep in the weak coupling region. The
corresponding values of the slow roll parameters are $\epsilon_{60} =
2\times10^{-5}$ and $\eta_{60} = -0.04$, which gives $A_S^2 =
5f_0^2/24\pi^4\Lambda^2$ and $n=0.91$. In order to satisfy the
constraint on the amplitude of perturbations~(\ref{ampli}), we require
$f_0=10^{-3}\Lambda$, which is a very natural value from the point of
view of the consistency of the theory.  In particular, these
parameters correspond to a vacuum mass scale of order
$V_0^{1/4}\simeq4\times10^{15}$\,GeV, very close to the GUT scale. For
other values of $\Lambda$ we find numerically the relation
$\log_{10}(f_0/\Lambda) = -4.5 + 1.54\,\log_{10} (M/\Lambda)$, which
is a very good fit in the range $1\leq M/\Lambda \leq 10^3$. For
$M>800\Lambda$, the soft breaking parameter $f_0$ becomes greater than
$\Lambda$, where our approximations breaks down, and we can no longer
trust our exact solution. Meanwhile the spectral tilt is essentially
invariant, $n=0.913 - 0.003\,\log_{10} (M/\Lambda)$, in the whole
range of $\Lambda$. It is therefore a concrete prediction of the
model. Surprisingly enough it precisely corresponds to the observed
value~(\ref{tilt}). This might change however when future satellite
missions will determine the spectral index $n$ with better than 1\%
accuracy~\cite{Kamion}.

There are also tensor (gravitational waves) metric perturbations in
this model, with amplitude $A_T^2 = 2(H/\pi M)^2 =
2f_0^2\Lambda^2/3\pi^4M^4$ and tilt $n_T = -2\epsilon$. The relative
contribution of tensor to scalar perturbations in the microwave
background on large scales can be parametrized~\cite{LL93} by
$T/S\simeq12.4\,\epsilon$. A very good fit to the ratio $T/S$ in this
model is given by $\,\log_{10}(T/S) = -2.6 - 0.91\,\log_{10}
(M/\Lambda)$, in the same range as above. Since $M\geq\Lambda$, we can
be sure that no significant contribution to the CMB temperature
anisotropies will arise from gravitational waves.

We have therefore found a new model of inflation, based on {\em exact}
expressions for the scalar potential of a softly broken N=2
supersymmetric SU(2) theory, to all orders in perturbations and with
all non-perturbative effects included. Inflation occurs along the weak
coupling Higgs region where the potential is essentially flat, and
ends when the gauge invariant scalar condensate enters the strong
coupling confining phase, where the monopole acquires a VEV, and
starts to oscillate around the minimum of the potential, reheating the
universe. A simple argument suggests that during reheating explosive
production of particles will occur in this model. The evolution
equation of a generic scalar (or vector) particle has the form of a
Mathieu equation and presents parametric resonance for certain values
of the parameters~\cite{KLS}. An efficient production of particles
occurs for large values of the ratio $q=g^2\Phi^2/4m^2$, where $g$
is the coupling between $\phi$ and the corresponding scalar field,
$\Phi$ is the amplitude of oscillations of $\phi$, and $m$ is its
mass. As the inflaton field oscillates around $\phi_{\rm min}$ it
couples strongly, $g\sim1$, to the other particles in the
supermultiplet since the minimum is in the strong coupling region. 
The amplitude of oscillations is of order the dynamical scale,
$\Phi\sim\Lambda$, while the masses of all particles (scalars,
fermions and vectors) are of order the supersymmetry breaking scale,
$m\sim f_0\ll\Lambda$~\cite{Luis2}.  This means that the
$q$-parameter is large, thus inducing strong parametric resonance and
explosive particle production~\cite{KLS}. These particles will
eventually decay into ordinary particles, reheating the universe. 

We are assuming throughout that we can embed this inflationary
scenario in a more general theory that contains two sectors, the
inflaton sector, which describes the soft breaking of N=2
supersymmetric SU(2) and is responsible for the observed flatness and
homogeneity of our universe, and a matter sector with the particle
content of the standard model, at a scale much below the inflaton
sector. The construction of more realistic scenarios remains to be
explored, in which the two sectors communicate via some messenger
sector. For example, one could consider this SU(2) as a subgroup of
the hidden $E_8$ of the heterotic string and the visible sector as a
subgroup of the other $E_8$. No-scale supergravity could then be used
as mediator of supersymmetry breaking from the strong coupling
inflaton sector to the weakly coupled visible sector. For the scales
of susy breaking considered above, $f_0\sim 10^{-5} M_{\rm P}$, we can
obtain a phenomenologically reasonable gravitino mass, $m_{3/2} \sim
f_0^3/M_{\rm P}^2 \sim 10$ TeV. This gravity-mediated supersymmetry
breaking scenario needs further study, but suggests that it is
possible in principle to do phenomenology with this novel inflationary
model.

\end{document}